# Traffic and Security using Randomized Dispersive Routes in Heterogeneous Sensor Network

P.Karunakaran[1] and Dr.C.Venkatesh[2]

[1]Department of Computer Science and Engineering, Erode Sengunthar Engineering College, Thudupathi, Erode
ctkaruna@gmail.com
[2]Dean, Faculty of Engineering, Erode Builder Educational Trust Institutions, Erode

## ABSTRACT

*Generally traffic and the sensor network security have many challenges in the transmission of data in the network. The existing schemes consider homogeneous sensor networks which have poor performance and scalability. Due to many-to-one traffic pattern, sensors may communicate with small portion of its neighbours. Key management is the critical process in sensor nodes to secure the data. Most existing schemes establish shared keys for all the sensors no matter whether they are communicating or not. Hence it leads to large storage overhead. Another problem in sensor network is compromised node attack and denial of service attack which occurs because of its wireless nature. Existing multi path routing algorithms are vulnerable to these attacks. So once an adversary acquires the routing algorithm, it can compute the same routes known to the source, and hence endanger all information sent over these routes. If an adversary performs node compromise attack, they can easily get the encryption/ decryption keys used by that node and hence they can intercept the information easily.*

*In this paper we are proposing a key management scheme which only establishes shared keys with their communicating neighbour and a mechanism to generate randomized multipath routes for secure transmission of data to the sink. Here we are adopting heterogeneous sensor networks and we are utilizing elliptic curve cryptography for efficient key management which is more efficient, scalable, and highly secure and reduces communication overhead. The routes generated by our mechanism are highly dispersive, energy efficient and making them quite capable of bypassing the back holes at low energy cost.*

## KEYWORDS

*Wireless Sensor Network, Non-Repetitive random propagation (NRRP) and Multi cat tree assisted random propagation (MTRP)*

## 1. INTRODUCTION

WIRELESS sensor networks have applications in many areas, such as military, homeland security, health care, environment, agriculture, manufacturing, and so on. In the past several years, sensor networks have been a very active research area. Most previous research efforts consider homogeneous sensor networks, where all sensor nodes have the same capabilities. However, a homogeneous ad hoc net-work suffers from poor fundamental limits and performance. Research has demonstrated its performance bottleneck both theoretically and through simulation experiments and test bed measurements. Several recent works studied Heterogeneous Sensor Networks (HSNs), where sensor nodes have different capabilities in terms of communication, computation, energy supply, storage space, reliability and other aspects.

Key management is an essential cryptographic primitive upon which other security primitives are built. Due to resource constraints, achieving such key agreement in wireless sensor networks is non-trivial. In Eschenauer and Gligor first present a key management scheme for sensor networks based on probabilistic key pre-distribution. Several other key pre-distribution schemes have been proposed. Probabilistic key pre-distribution is a promising scheme for key management in sensor networks. To ensure such a scheme works well, the probability that each sensor shares at least one key with a neighbour sensor (referred to as key-sharing probability) should be high.

The above discussion shows that many existing key management schemes require a large storage space for key pre-distribution and are not suitable for small sensor nodes. Of the various possible security threats that may be experienced by a wireless sensor network (WSN), in this paper we are specifically interested in combating two types of attacks: the compromised-node (CN) attack and the denial-of-service (DOS) attack. The CN attack refers to the situation when an adversary physically compromises a subset of nodes to eavesdrop information, whereas in the DOS attack, the adversary interferes with the normal operation of the WSN by actively disrupting, changing, or even destroying the functionality of a subset of nodes in the system. These two attacks are similar in the sense that they both generate *black holes*: areas within which the adversary can either passively intercept or actively block information delivery. Due to the unattended nature of WSNs, adversaries can easily produce such black holes. Severe CN and DOS attacks can disrupt normal data delivery between sensor nodes and the sink, or even partition the topology. A conventional cryptography-based security method cannot alone provide satisfactory solutions to these problems. This is because, by definition, once a node is compromised, the adversary can always acquire the encryption/decryption keys of that node, and thus can intercept any information passed through it. At the same time, an adversary can always perform certain form of DOS attack (e.g., jamming) even if it does not have any knowledge of the crypto-system used in the WSN.

One remedial solution to these attacks is to exploit the network's routing functionality. Specifically, if the locations of the black holes formed by the compromised (or jammed) nodes are known a priori, then information can be delivered over paths that circumvent (bypass) these holes, whenever possible. In practice, due to the difficulty of acquiring such location information, the above idea is implemented in a probabilistic manner, typically through a two-step process: secret sharing and multi-path routing.

We argue that three security problems exist in the counter-attack approach. First, this approach is no longer valid if the adversary can *selectively* compromise or jam nodes. This is because the route computation in the above multi-path routing algorithms is deterministic in the sense that for a fixed topology, a fixed set of routes is always computed by the routing algorithm for given source and destination. Therefore, even if the shares can be distributed over different routes, overall they are always delivered over the same set of routes that are computable by the algorithm. As a result, once the routing algorithm becomes open to the adversary (this can be done, e.g., through a memory interrogation of the compromised nodes), the adversary can by itself compute the set of routes for any given source and destination. Then the adversary can pinpoint to one particular node in each route and compromise (or jam) these nodes. Such an attack can intercept all shares of the information, rendering the above counter-attack approaches ineffective. Second, as pointed out in, actually very few node-disjoint routes can be found when node density is moderate and source and destination nodes are several hops apart. For example, for a node degree of 8, on average only two node-disjoint routes can be found between a source and a destination that are at least 7 hops apart. There is also a 30% possibility that no node-disjoint paths can be found between the source and the destination. The lack of enough routes significantly undermines the security performance of this multi-path approach. Last, even worse, because the set of routes is computed under certain constraints, the routes may not be spatially dispersive enough to circumvent a moderate-sized black hole.

In this paper, we present an efficient key management scheme and a randomized multipath routing algorithm that only needs small storage space less energy consumption. Contribution of this paper is of three folds. First, we utilize the C-neighbour concept and a key management scheme for power full sensors. Second, establishing keys among sensors using ECC public key cryptosystem. Third, developing distributed multipath routing algorithms based on the information available to the sensors. The schemes proposed are Non-Repetitive random propagation (NRRP) and Multi cat tree assisted random propagation (MTRP). The rest of the paper is organized as follows. Section II describes the proposed key management scheme. Section III describes the proposed multi path routing algorithm. Section IV describes the simulation results of the proposed schemes. Section V describes the conclusion.

## 2. PROPOSED KEY MANAGEMENT SCHEME

### 2.1 The Cluster Formation

After sensor deployment, clusters are formed in an HSN. We have designed an efficient clustering scheme for HSNs in. Because of the page limit, we will not describe the details of the clustering scheme here. For the simplicity of discussion, assume that each H-sensor can communicate directly with its neighbour H-sensors (if not, then relay via L-sensors can be used). All H-sensors form a backbone in an HSN. After cluster formation, an HSN is divided into multiple clusters, where H-sensors serve as the cluster heads. An illustration of the cluster formation is shown in Figure 1, where the small squares are L-sensors, large rectangular nodes are H-sensors, and the large square at the bottom-left corner is the sink.

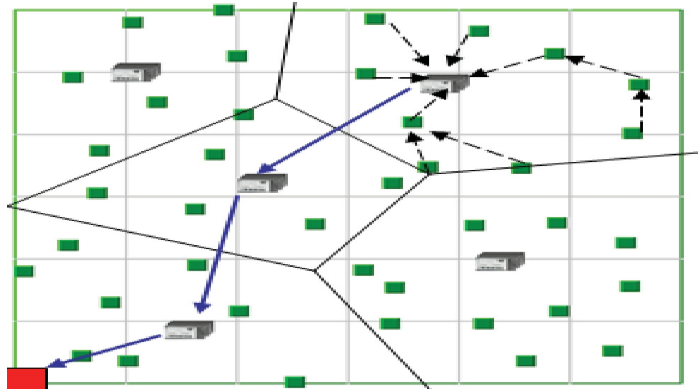

Figure 1. Cluster Formation in HSN

### 2.2 Distributed Key Establishment

The key setup can also be done in a distributed way. In the distributed key establishment, each L-sensor is pre-loaded with a pair of ECC keys - a private key and a public key. When an L-sensor (denoted as $u$) sends its locations information to its cluster head H, $u$ computes a Message Authentication Code (MAC) over the message by using $u$'s private key, and the MAC is appended to message. When H receives the message, H can verify the MAC and then authenticate $u$'s identify, by using $u$'s public key. Then H generates a certificate (denoted as CA$u$) for $u$'s public key by using H's private key.

After determining the routing tree structure in a cluster, the cluster head H disseminates the tree structure (i.e., parent-child relationship) and the corresponding public key certificate to each L-sensor. The public key certificates are signed by H's private key, and can be verified by every L-sensor, since each L-sensor is preloaded with H's public key. A public key certificate proves the authenticity of a public key and further proves the identity of one L-sensor to another L-sensor.

If two L-sensors are parent and child in the routing tree, then they are $c$-neighbours of each other, and they will set up a shared key by themselves. For each pair of $c$-neighbours, the sensor with smaller node ID initiates the key establishment process. For example, suppose that L-sensor $u$ and $v$ are $c$-neighbours and $u$ has a smaller ID than $v$. The process is presented below:

1) Node $u$ sends its public key $K_u^U = I_u P$ to $v$.

2) Node $v$ sends its public key $K_v^U = I_v P$ to $u$.

3) Node $u$ generates the shared key by multiplying its private key $I_u$ with $v$'s public key - $K_v^U$, i.e., $K_{u,v} = K_u^R K_v^U = I_u I_v P$; similarly, $v$ generates the shared key

$$= K_v^R K_u^U = I_u I_v P .$$

After the above process, nodes $u$ and $v$ share a common key and they can start secure communications. To reduce the computation overhead, symmetric encryption algorithms are used among L-sensors. Note that in the distributed key establishment scheme, the assumption of having tamper-resistant hardware in H-sensors can be removed.

# 3. PROPOSED RANDOMIZED MULTIPATH ALGORITHM

## 3.1. Overview

As illustrated in Figure 2, we consider a 3-phase approach for secure information delivery in a WSN: secret sharing of information, randomized propagation of each information share, and normal routing (e.g., min-hop routing) toward the sink. More specifically, when a sensor node wants to send a packet to the sink, it first breaks the packet into $M$ shares according to a $(T; M)$-threshold secret sharing algorithm, e.g., the Shamir's algorithm. Each share is then transmitted to some randomly picked neighbor. That neighbor will continue to relay the share it has received to other randomly picked neighbors, and so on. In each information share, there is a TTL field, whose initial value is set by the source node to control the total number of randomized relays. After each relay, the TTL field is reduced by 1. When the TTL count reaches 0, the final node receiving this share stops the random propagation phase and begins to route this share towards the sink using normal single-path routing. Once the sink collects at least $T$ shares, it can inversely compute the original information. No information can be recovered from less than $T$ shares.

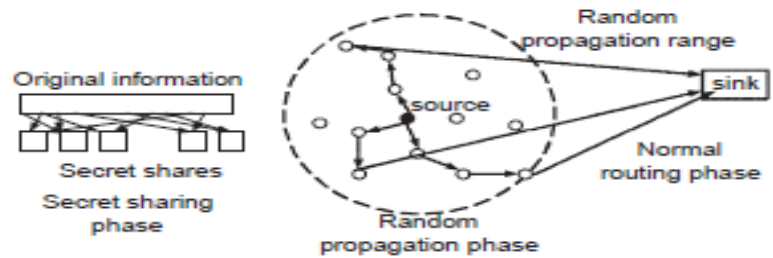

Figure 2. Randomized Dispersive routes

Because routes are randomly generated, there is no guarantee that different routes are still node-disjoint. However, the algorithm should ensure that the randomly generated routes are as dispersive as possible, i.e., different routes are geographically separated as far as possible such that they have high likelihood of not simultaneously passing through a black hole. Considering the stringent requirement on energy consumptions in WSNs, the major challenge in our design is to generate highly dispersive random routes at low energy cost.As explained later, such a challenge is not trivial. A naive algorithm of generating random routes, such as Wanderer scheme (a pure random-walk algorithm), only leads to long paths (containing many hops, and therefore, consuming much energy) without achieving good dispersive ness. Due to security considerations, we also require that the route computation be implemented in a distributed way, such that the final route represents the aggregate decision of all the nodes participating in route selection. As a result, a small number of colluding/compromised nodes cannot dominate the selection result. In addition, for efficiency purposes, we also require that the randomized route selection algorithm only incurs a small amount of communication overhead. Needless to say, the random propagation phase is the key component that dictates the security and energy performance of the entire mechanism.

## 3.2. Random propagation of Information Shares

To diversify routes, an ideal random propagation algorithm propagates information shares as dispersive as possible. Typically, this means propagating the share farther from its source. At the same time, it is highly desirable to have an energy-efficient propagation, which calls for limiting the number of randomly propagated hops. The challenge here lies in the random and distributed nature of the propagation: a share may be sent one-hop farther from its source in a given step, but may be sent back closer to the source in the next step, wasting both steps from the security's point of view. To tackle this issue, some control needs to be imposed on the random propagation process to ensure that in each step the share is more likely to be forwarded outwards from the source. We develop four distributed random propagation mechanisms, which approach this goal in various degrees.

***Non-repetitive Random Propagation***: NRRP is based on PRP, but it improves the propagation efficiency by recording all the nodes that the propagation has traversed so far. More specifically, NRRP adds a "node-in-route" (NIR) field to the header of each share. Initially, this field is empty. Starting from the source node, whenever a node propagates the share to the next hop, the id of the up-stream node is appended to the share's NIR field. Nodes included in NIR are excluded from the random pick of the next hop of propagation. This non-repetitive propagation guarantees that the share will be relayed to a different node in each step of random propagation, leading to better propagation efficiency.

***Multicast Tree-assisted Random Propagation***: The MTRP scheme aims at actively improving the energy efficiency of random propagation while preserving the dispersiveness of DRP. The basic idea comes from the following observation of Figure 2: Among the 3 different routes taken by the shares, the route on the bottom right is the most energy efficient because it has the shortest end-to-end path. So, in order to improve energy efficiency, the shares should be best propagated in the direction of the sink. In other words, their propagation should be restricted to the right half of the circle in Figure 2.

Conventionally, directional routing requires location information of both the source and the destination nodes, and sometimes the intermediate nodes. Examples of this type of location-based routing are GPSR (Greedy Perimeter Stateless Routing) and LAR (Location-Aided Routing). Location in-formation mainly relies on GPS in each node, or on some distributed localization algorithms. The high cost and the low accuracy of localization are the main drawbacks of these two methods, respectively.

MTRP involves directionality in its propagation process without needing location information. More specifically, after the deployment of the WSN, MTRP requires that the sink constructs a multicast tree from itself to every node in the network. Such a tree-construction operation is not unusual in existing protocols, and is typically conducted via flooding a "hello" message from the sink to every node. Once this multicast tree is constructed, a node knows its distance (in number of hops) to the sink and the id of its parent node. We assume that each entry in the neighbor list maintained by a node has a field recording the number of hops to the sink from the corresponding neighbor. Under MTRP, the header of each share contains two additional fields: $\max_{hop}$ and $\min_{hop}$. The values of these two parameters are set by the source to $\max_{hop} = n_s + \alpha_1$ and $\min_{hop} = n_s - \alpha_2$, where $n_s$ is the hop count from the source to the sink, and $\alpha_1$ and $\alpha_2$ are nonnegative integers with $\alpha_1 \leq \alpha_2$. The parameter $\alpha_1$ controls the limit that a share can be propagated away from the sink, i.e., to the left half of the circle in Figure 1. The parameter $\alpha_2$ controls the propagation area toward the sink, i.e., the right half of the circle. A small $\alpha_2$ makes the propagation of a share be dispersed away from the centre line connecting the source and the link and forces them to take the side path, leading to better dispersion.

Before a node begins to pick the next relaying node from its neighbor list, it first filters out neighbors that are in the LHNL, just as in the case of DRP. Next, it filters out nodes that have a hop count to the sink greater than $\max_{hop}$ or smaller than $\min_{hop}$. The next relaying node will be randomly drawn from the remaining neighbors. In case the set of remaining nodes after the first step is empty, the second step will be directly applied to the entire set of neighbors.

## 4. SIMULATION RESULTS

### 4.1. Performance Evaluation of Key Management Scheme

In this Section, we present the performance evaluation results of the ECC-based key management scheme (referred to as the ECC scheme below). The key pre-distribution scheme proposed by Eschenauer and Gligor [6] is used for comparison, and it is referred to as the E-G scheme.

Table 1: Performance Evaluation

| Schemes | Storage Space | Energy Consumption | Security |
|---|---|---|---|
| Basic Scheme | **m (M + N)** *where m depends on the key pool size P* | With the key pool size p=10000, the probability of sharing is 90% | When more keys are pre-loaded in a sensor, the compromising probability is high. (i.e) less resilient to node compromise attack |
| Distributed Key Setup | **3M + 2N** *where M is the number of H Sensors and N is the number of L Sensors* | Establishes keys only between the c-neighbors | Compromising probability is always zero. Since each L-sensor uses a distinctive public/private key pair. |

### 4.2. Evaluation of Multi Path Routing Algorithm

We first fix the location of the source node at (¡50; 0). In Figures 3 and 4, we plot the packet interception probability as a function of the TTL value (*N*) and the number of shares (*M*) that each packet is broken into, respectively. These figures show that increasing *N* and *M* helps reduce the packet interception probability for all proposed schemes. However, for a sufficiently large *N*, (e.g., *N* = 20 in Figure 3), the interception probability does not change much with a further increase in *N*. This is because the random propagation process has reached steady state. It can also be observed that, in all cases, the packet interception probabilities under the DRP, NRRP, and MTRP schemes are much smaller than that of the baseline PRP scheme, because their random propagations are more efficient. In addition, when $N$ and $M$ are large, all four randomized algorithms achieve smaller packet interception probabilities than the deterministic H-SPREAD scheme. In many cases, the gap is more than one order of magnitude. We plot the packet interception probability as a function of the size of the black hole in Figure 5. It is clear that the interception probability increases with $R_e$.

In Figure 6 we study the impact of node connectivity. The number of nodes is changed from 1000 to 3000, corresponding to changing the average node connectivity degree from 8 to 24. It can be observed that, in general, the packet interception probabilities of the four proposed schemes do not change significantly with node connectivity. Such insensitivity to node connectivity/density is because the packet interception probability is mainly decided by how dispersive the shares can be geographically after random propagation. As long as nodes are uniformly distributed, a change in node density does not impact the geographic distribution of the shares after random propagation.

In Figure 7, we slide the x-coordinate of the source node along the line $y = 0$ to evaluate the packet interception probabilities at different source locations in the network. A segmented trend can be observed: when the source is far away from the black hole, shares are mainly intercepted during the normal routing phase. Note that during the normal routing phase, all paths start to converge geographically to the sink (see Figure 3). As a result, the closer the source is to the black hole, the less convergent the paths will be at the black hole, so the lower interception probability. However, when the source is close to the black hole, i.e., $x$ ₁ 0, the trend in the interception probability is reversed. This is because more and more shares are intercepted during the propagation phase. When $x = 50$, which corresponds to the scenario where the source is placed right at the center of the black hole, the interception probabilities reach their maximum value. After that, they decrease quickly as the source gets farther away from the black hole. In all segments, the packet interception probabilities of the DRP, NRRP, and MTRP schemes are smaller than that of H-SPREAD.

We evaluate the average number of hops of the end-to-end route as a function of the TTL value in Figure 8. It can be observed that the hop-count under PRP, DRP, and NRRP increases linearly with $N$, while the hop-count under MTRP only increases slowly with $N$. The TTL value does not play a role in the H-SPREAD scheme. Under large $N$, e.g., when $N = 25$, the randomized algorithm achieves better security performance than H-SPREAD. However, the hop-count of H-SPREAD is about 1∕3 of that of PRP, DRP, and NRRP, and about 1∕2 of that of MTRP scheme. The relatively large hop-count in the randomized algorithms is the cost for stronger capability of bypassing black holes.

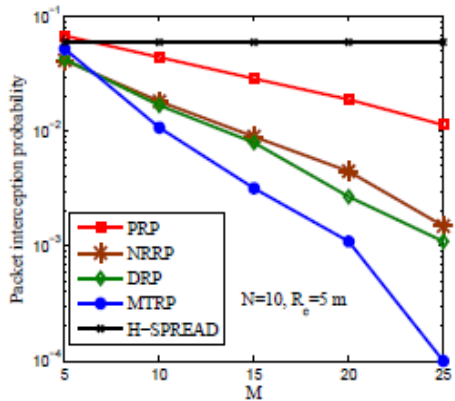

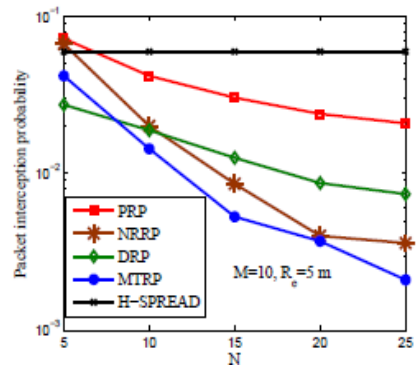

Figure3. Packet interception probability vs N        Figure 4. Packet interception probability vs M

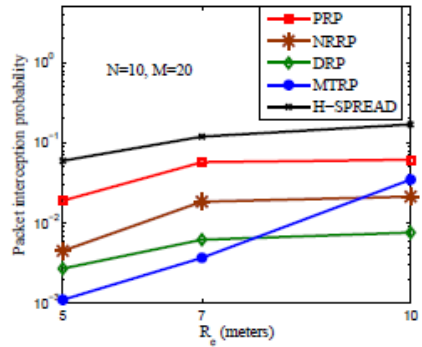

Figure 5: Packet Interception probability vs $R_e$.

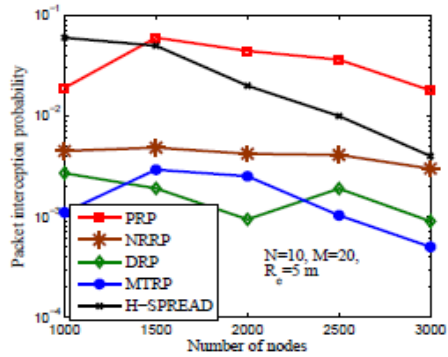

Figure 6: Packet interception probability vs number of nodes

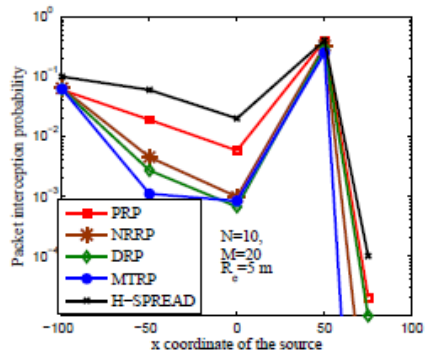

Figure 7: Packet interception probability vs different source location

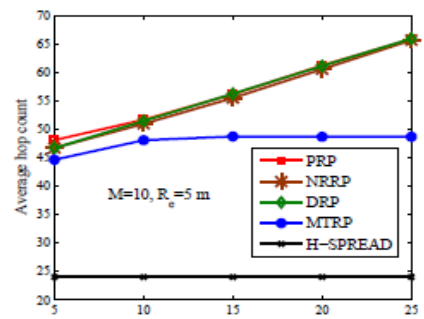

Figure 8: Hop count vs N

## 5. CONCLUSION

 The proposed key management scheme utilizes the fact that a sensor only communicates with a small portion of its neighbors and thus greatly reduces the communication and computation overheads of key setup. A public key algorithm – Elliptic Curve Cryptography (ECC) is used to further improve the key management scheme. The scheme only pre-loads a few keys on each L-sensor and thus significantly reduces sensor storage requirement. Our performance evaluation and security analysis showed that the routing-driven, ECC-based key management scheme can significantly reduce communication overhead, sensor storage requirement and energy consumption while achieving better security than a popular key management scheme for sensor networks.

The simulation results have shown the effectiveness of randomized dispersive routing in combating CN and DOS attacks. By appropriately setting the secret sharing and propagation parameters, the packet interception probability can easily be reduced by the proposed algorithms to as low as $10^{/3}$, which is at least one order of magnitude smaller than approaches that use deterministic node-disjoint multi-path routing. At the same time, we have also verified that this improved security performance comes at a reasonable cost of energy.

**Authors**

P.Karunakaran is working as Lecturer at Erode Sengunthar Engineering College, Thudupathi, Tamilnadu, India. He has received M.Sc., M.Phil. M.E Degree in Computer Science and Engineering, currently pursuing Ph.D at Anna University of Technology, Coimbatore. Totally he has more than five years of experience in teaching. His research interest includes Network Routing, Routing Architecture and Network Security.

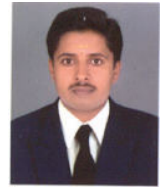

Dr. C. Venkatesh, graduated in ECE from Kongu Engineering College in the year 1988, obtained his master degree in Applied Electronics from Coimbatore Institute of Technology, Coimbatore in the year 1990. He was awarded Ph D in ECE from Jawaharlal Nehru Technological University, Hyderabad in 2007. He has a credit of two decade of experience which includes around 3 years in industry. He has 16 years of teaching experience during tenure he was awarded **Best Teacher Award** twice. He was the founder Principal of Surya Engineering College, Erode. He is guiding 10 Ph.D., research scholars. He is a Member of IEEE, CSI, ISTE and Fellow IETE. He has Published 13 papers in International and National Level Journals and 50 Papers in International and National Level conferences. His area of interest includes Soft Computing, Sensor Networks and communication.

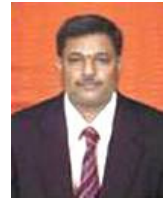